\documentclass[preprint,showpacs,preprintnumbers,amsmath,amssymb,superscriptaddress,prl]{revtex4}

\usepackage{graphicx}
\usepackage{dcolumn}
\usepackage{bm}

\begin{document}

\title{Picosecond spin Seebeck effect}

\author{Johannes Kimling}
\email{kimling@illinois.edu}
\affiliation{Department of Materials Science and Engineering and Materials Research Laboratory, University of Illinois, Urbana, Illinois 61801, USA}

\author{Gyung-Min Choi}
\affiliation{Center for Spintronics, Korea Institute of Science and Technology, Seoul 136-791, Korea}

\author{Jack T. Brangham}
\affiliation{Department of Physics, The Ohio State University, 191 West Woodruff Avenue, Columbus, Ohio 43210, USA}

\author{Tristan Matalla-Wagner}
\affiliation{Center for Spinelectronic Materials and Devices, Department of Physics, Bielefeld University, Universit\"atsstrasse 25, 33615 Bielefeld, Germany}

\author{Torsten Huebner}
\affiliation{Center for Spinelectronic Materials and Devices, Department of Physics, Bielefeld University, Universit\"atsstrasse 25, 33615 Bielefeld, Germany}

\author{Timo Kuschel}
\affiliation{Center for Spinelectronic Materials and Devices, Department of Physics, Bielefeld University, Universit\"atsstrasse 25, 33615 Bielefeld, Germany}
\affiliation{Physics of Nanodevices, Zernike Institute for Advanced Materials, University of Groningen, Nijenborgh 4, 9747 AG Groningen, The Netherlands}

\author{Fengyuan Yang}
\affiliation{Department of Physics, The Ohio State University, 191 West Woodruff Avenue, Columbus, Ohio 43210, USA}

\author{David G. Cahill}
\affiliation{Department of Materials Science and Engineering and Materials Research Laboratory, University of Illinois, Urbana, Illinois 61801, USA}
\date{\today}

\pacs{Valid PACS appear here}

\begin{abstract}
We report time-resolved magneto-optic Kerr effect measurements of the longitudinal spin Seebeck effect driven by an interfacial temperature difference between itinerant electrons and magnons. The measured time-evolution of spin accumulation induced by laser-excitation indicates transfer of angular momentum across Au/Y$_3$Fe$_5$O$_{12}$ and Cu/Y$_3$Fe$_5$O$_{12}$ interfaces on a picosecond time-scale. The product of spin-mixing conductance and interfacial spin Seebeck coefficient determined is of the order of $10^8$~A~m$^{-2}$~K$^{-1}$.
\end{abstract}

\maketitle

\par \textit{Introduction.-} The longitudinal spin Seebeck effect (LSSE) describes the appearance of a spin current through the interface between a normal metal and a magnetic insulator if a heat current is flowing perpendicular through that interface \cite{Uchida2010}. Possible application of the LSSE is envisaged for information technologies as well as for new routes for conversion of heat into electric energy \cite{Hoffmann2015}.

The LSSE involves transfer of angular momentum across a metal/insulator interface. Interfacial exchange interaction provides a possible coupling mechanism between itinerant electrons and localized electrons across the interface \cite{Zhang2012,Bender2012}. Based on this coupling mechanism and provided that the insulator supports magnons, itinerant electrons scattering off the interface can create or annihilate magnons, thus allowing for interconversion of spin current and magnon current. 

\par LSSE theories consider thermally excited spin currents in both directions across a metal/insulator interface: A spin current from insulator to metal driven by a thermal spin pumping mechanism, and a spin current from metal to insulator driven by random spin transfer torques \cite{Xiao2010,Adachi2013,Bender2015}. In equilibrium, these opposite currents are equal. Application of a temperature gradient creates an imbalance of the thermally excited spin currents. The net spin current is predicted to be proportional to the interfacial temperature difference between electrons and magnons. In addition to this interfacial LSSE, a temperature gradient in the bulk of the insulator can drive a magnon current that results in accumulation (depletion) of magnons near the interface enhancing (reducing) the spin current from insulator to metal \cite{Brechet2013,Hoffman2013,Ritzmann2014,Rezende2014}. To date, isolation of interfacial LSSE from bulk LSSE has not been achieved experimentally.
 
\par Prior LSSE measurements are based on the inverse spin Hall effect (ISHE): the voltage signal measured is assumed to be caused by a spin current that has been converted into a transverse charge current. The symmetry of the resulting voltage signal with respect to the applied magnetic field direction is used as an indication of the ISHE. The ISHE-based measurement of spin currents is also used in related experiments on ferromagnetic resonance spin pumping and long-distance magnon currents \cite{Du2015,Cornelissen2015}.

\par ISHE-based LSSE measurements have been reported for various insulators, e.g., ferrimagnetic garnets such as Y$_3$Fe$_5$O$_{12}$ \cite{Uchida2013}, Bi-substituted Y$_3$Fe$_5$O$_{12}$ \cite{Siegel2014}, and Gd$_3$Fe$_5$O$_{12}$ \cite{Geprags2016}, ferrimagnetic ferrites such as NiFe$_2$O$_4$ \cite{Meier2013,Meier2015}, CoFe$_2$O$_4$ \cite{Niizeki2015,Guo2016}, and Fe$_3$O$_4$ \cite{Ramos2015}, as well as paramagnetic Gd$_3$Ga$_5$O$_{12}$ \cite{Wu2015} and antiferromagnetic Cr$_2$O$_3$ \cite{Seki2015} or MnF$_2$ \cite{Wu2016}. The experiments are typically reported as observations of the LSSE. However, ISHE-based LSSE measurements are susceptible to unwanted voltage sources, e.g., proximity Nernst effects \cite{Huang2010} and conventional Seebeck effect driven by thermal Hall heat current in the ferromagnetic layer \cite{Wegrowe2014}. Hence, independent LSSE measurements that are not based on the ISHE are required to corroborate the spin current hypothesis of the LSSE.

\par To date, time-resolved ISHE-based LSSE measurements have achieved a time-resolution of the order of 10-100~ns \cite{Agrawal2014,Roschewsky2014}. Agrawal~\textit{et~al.} investigate $\mu$m-thick YIG layers and report that the time-scale of the LSSE is determined by the rise-time of the temperature gradient in the YIG layer ($\sim$300~ns). They conclude that the LSSE is predominantly a bulk effect caused by magnon diffusion along the temperature gradient in the YIG layer. Based on this interpretation, they estimate a magnon diffusion length of $\sim$500~nm. ISHE-based LSSE measurements as function of YIG thickness support the dominant role of the bulk LSSE for YIG-thicknesses of the order of the magnon diffusion length \cite{Kehlberger2015}. Roschewsky~\textit{et~al.} investigate YIG layers with thicknesses of $\sim$50~nm and report a constant LSSE signal for heating frequencies up to 30~MHz. They conclude that the characteristic time-scale of the LSSE was shorter than 5~ns.

\par Here, we present a LSSE experiment that is based on the time-resolved magneto-optic Kerr effect (TR-MOKE) and provides sub-picosecond time resolution. Our experiment is not susceptible to spurious effects that plague ISHE-based LSSE measurements. Taking advantage of the picosecond time-scale, our experiment involves sizable temperature differences across the NM/YIG interface, of the order of 10 to 100~K. This allows us to selectively probe the interfacial LSSE.

\par \textit{Experiment and model.-} The samples are normal metal (NM)/YIG bilayers on Gd$_3$Ga$_5$O$_{12}$ (GGG) substrates. Since we measure spin accumulation in the NM layer, we use Au and Cu as NM materials with long spin-relaxation times (one order of magnitude longer compared to Pt) \cite{Choi2014}. 

\par Sample~I-III were grown at Ohio State University by off-axis sputtering and in-situ deposition of Au for sample~I and sample~II and ex-situ deposition of Cu for sample~III. Sample IV-VI were grown in collaboration between University of Alabama and University of Bielefeld, Germany. The YIG of samples IV-VI was deposited by pulsed-laser deposition. For sample~IV and sample~V, Au was ex-situ sputtered on as-grown YIG/GGG; for sample~VI, Cu was sputtered in-situ after vacuum annealing of YIG/GGG at 200$^\circ$C and $4.6\times10^{-9}$~mbar for 1~h. The measurements were done at University of Illinois. 

\par We excite the NM layer with a train of optical pulses at a repetition rate of 80~MHz and absorbed fluences of the order of 1~J~m$^{-2}$ \cite{Suppl}. The absorbed laser energy is initially deposited in the heat capacity of electrons and transferred to the phonon heat capacity via electron-phonon scattering. To describe this heat transfer problem, we use a two-temperature model of electrons and phonons,
\begin{eqnarray}
\label{Eq1}
C_{\rm e}\frac{\partial T_{\rm e}}{\partial t} - \Lambda_{\rm e}\frac{\partial^2T_{\rm e}}{\partial x^2} &=& g_{\rm ep}(T_{\rm p}-T_{\rm e}) + p(z,t),\\
\label{Eq2}
C_{\rm p}\frac{\partial T_{\rm p}}{\partial t} - \Lambda_{\rm p}\frac{\partial^2T_{\rm p}}{\partial x^2} &=& g_{\rm ep}(T_{\rm e}-T_{\rm p}),
\end{eqnarray}
where $C$ denotes volumetric heat capacity, $\Lambda$ denotes thermal conductivity, and $g_{\rm ep}$ denotes coupling parameter between electrons (e) and phonons (p), and $p(z,t)$ is the optical absorption profile determined using an optical transfer matrix method \cite{Suppl}. We assume that the electronic heat capacity is proportional to the electron temperature, $C_{\rm e} = \gamma_{\rm e} T_{\rm e}$, where $\gamma_{\rm e}$ is the electronic heat capacity coefficient.  

\par
\begin{figure}[h!]
	\centering
		\includegraphics[trim = 0mm 0mm 0mm 0mm,clip,width=8.6cm]{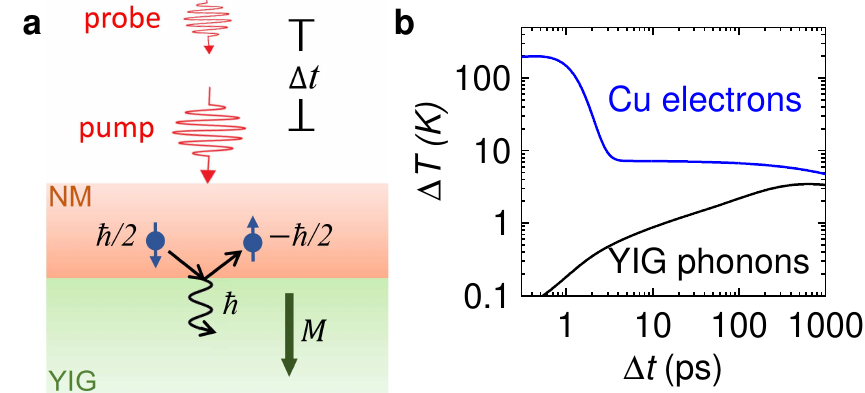}
	\caption{(Color online). \textbf{Conceptual diagram and temperature transients.} (a) Absorption of a pump laser pulse of picosecond duration generates a temperature difference between NM electrons and YIG magnons. Interfacial coupling between electrons and magnons induces spin accumulation in NM, which is probed by time-delayed probe laser pulses. (b) Exemplary temperature transients of Cu electrons and YIG phonons calculated using the two-temperature model, Eqs.~(\ref{Eq1}) and (\ref{Eq2}). }
	\label{Fig1}
\end{figure}

 The temperature excursion of electrons is of the order of 100~K during laser excitation (compare Fig.~\ref{Fig1}). After thermalization of electrons and phonons in the NM layer, the finite thermal conductance of the NM/YIG interface maintains a temperature-difference between electrons and YIG-phonons of the order of 10~K for approximately 100~ps. Energy exchange across the NM/YIG interface is dominated by phonons. Energy transfer to YIG-magnons can occur via direct coupling of electrons and magnons across the NM/YIG-interface and through phonon-magnon coupling of YIG. 

\par SSE theories predict that the temperature difference between YIG magnons and NM electrons drives a spin current across the NM/YIG interface \cite{Xiao2010,Adachi2013,Bender2015}:
\begin{equation}
\label{Eq3}
j_{\rm S} = g_{\uparrow\downarrow}\frac{e^2}{h}S_{\rm S} (T_{\rm e}-T_{\rm m}),
\end{equation}
where $g_{\uparrow\downarrow}$ is the real part of the spin-mixing conductance per conductance quantum $\frac{e^2}{h}$ and $S_{\rm S}$ is the interfacial spin Seebeck coefficient. Since magnon heat capacity and phonon-magnon coupling parameter of YIG are unknown, we approximate the magnon temperature by the average phonon temperature of the YIG layer determined from the two-temperature model. In the results section below, we provide arguments that support the validity of this approximation. 

\par During the pump-probe measurements, a magnetic field of $\sim$0.4~T perpendicular to the sample plane rotates the YIG magnetization out-of-plane. If a significant amount of spin accumulation is generated in the NM layer, the resulting non-equilibrium magnetization rotates the polarization of light upon reflection \cite{Choi2014}. We probe this polar Kerr effect with a train of sub-ps optical pulses at the same repetition rate of 80~MHz and a lower absorbed fluence of approximately 0.03~J m$^{-2}$. By varying the time delay between successive pump and probe pulses, we track rise and decay of spin accumulation subsequent to laser excitation \cite{Choi2015}. To determine zero time delay and temporal heating profile, we use a GaP photodiode at the sample location, which measures the temporal profile of correlated pump and probe pulses. The magnitude of the polar Kerr signal for a given amount of spin accumulation is determined by the strength of spin-orbit coupling \cite{Choi2014}. We use conversion factors between polar Kerr rotation and spin accumulation estimated in prior works ($4.8\times10^{-8}$~rad~m~A$^{-1}$ for Au \cite{Choi2014} and $0.9\times10^{-8}$~rad~m~A$^{-1}$ for Cu \cite{Choi2015}). A description of the experimental setup can be found in the Supplemental Material \cite{Suppl} including Refs.~\onlinecite{Schreier2013,Touloukian1971,Tari2003,Choi2014,Choi2015,Scott1974,Wood1990}.

\par
\begin{figure}[h!]
	\centering
		\includegraphics[trim = 0mm 0mm 0mm 0mm,clip,width=17.8cm]{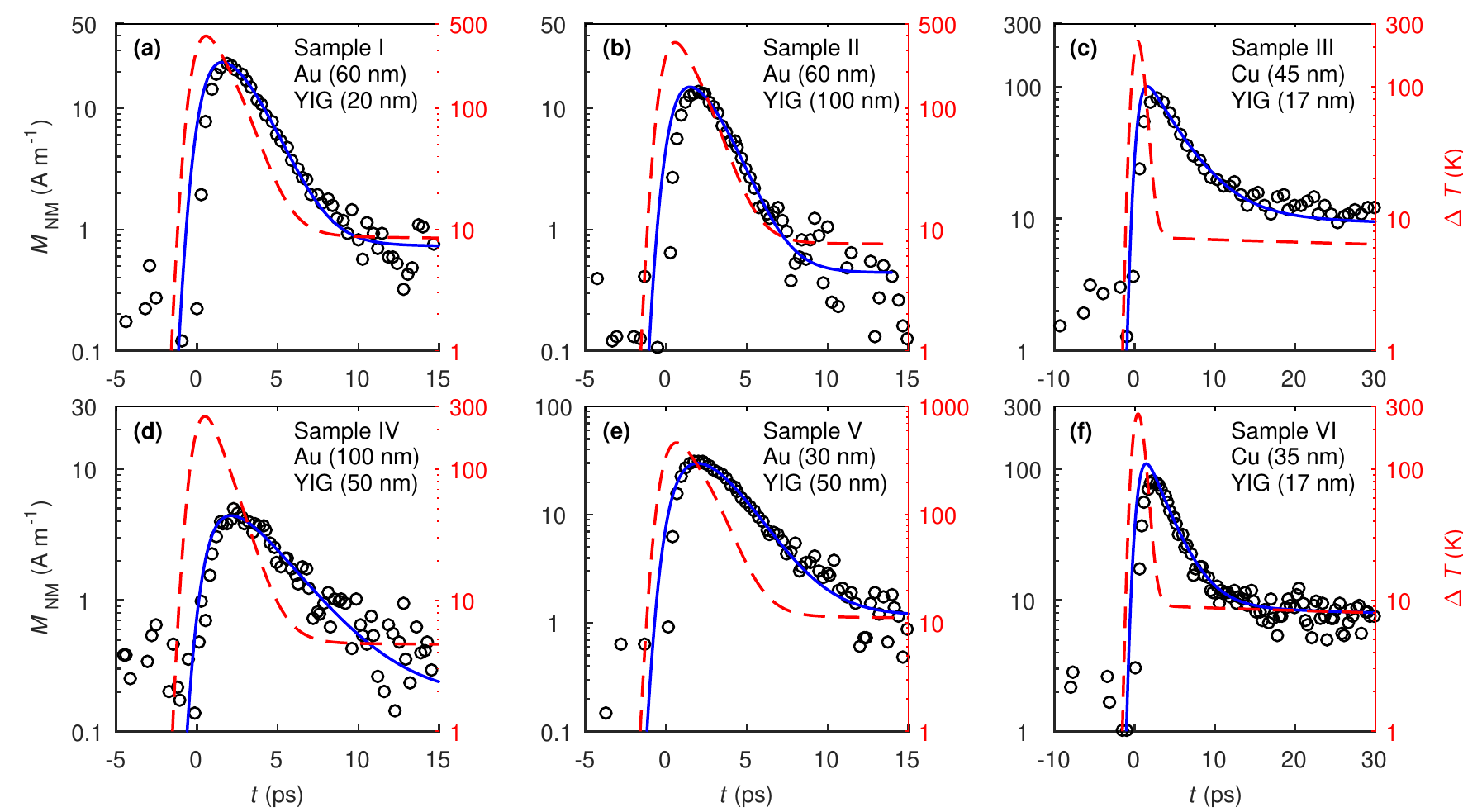}
	\caption{(Color online). \textbf{Thermally-induced spin accumulation in Au and Cu.} TR-MOKE data (circles) measured on Au (Cu)/YIG/GGG samples of different Au (Cu) thicknesses as indicated in the figure. Solid lines show fit curves obtained using the spin-diffusion model, Eqs.~(\ref{Eq3}) and (\ref{Eq4}). Dashed lines show temperature excursion of electrons calculated using the two-temperature model, Eqs.~(\ref{Eq1}) and (\ref{Eq2}).}
	\label{Fig2}
\end{figure}

 To describe spin accumulation in the NM layer, we consider the time-dependent spin diffusion equation
\begin{equation}
\label{Eq4}
\frac{\partial\zeta_{\rm S}}{\partial t}-D\frac{\partial\zeta_{\rm S}}{\partial x^2} = \frac{\zeta_{\rm S}}{\tau_{\rm S}},
\end{equation}
and connect the spin current in equation~(\ref{Eq3}) with the spin diffusion current $j_{\rm S} = \frac{\sigma}{2e}\frac{\partial\zeta_{\rm S}}{\partial x}$
at the NM/YIG interface. In the above equation, $\zeta_{\rm S} = \zeta_{\uparrow}-\zeta_{\downarrow}$ is the difference of the chemical potentials of up- and down-spins, $\sigma$ is the electrical conductivity, $D = \sigma/[e^2N(E_{\rm F})]$ is the diffusion constant of electrons, where $N(E_{\rm F})$ is the electronic density of states at the Fermi energy, and $\tau_{\rm S}$ is the spin relaxation time. We fit the solution of the spin diffusion model to the measurement data using $\tau_{\rm S}$ and the product of spin-mixing conductance and spin Seebeck coefficient, $\alpha\equiv g_{\uparrow\downarrow}\frac{e^2}{h}S_{\rm S}$, as free parameters (compare Fig.~\ref{Fig2}). Due to the large diffusion constant of electrons in Au and Cu, the spin accumulation near the NM surface does not vary significantly within the optical absorption depth. Therefore, we assume that the probe measures the spin accumulation at the NM surface. The sensitivity of spin accumulation to $\alpha$ is a constant; the sensitivity of spin accumulation to $\tau_{\rm S}$ peaks shortly after laser-excitation, when the temperature excursion of electrons falls back to the phonon temperature \cite{Suppl}. 

\par \textit{Results.-} The measurement signal rises during laser-excitation and decays to a plateau a few picoseconds after laser-excitation [open circles in Fig.~\ref{Fig2}~(a)-(f)]. The remaining measurement signal decays slowly with the interfacial temperature difference for approximately 1~nanosecond \cite{Suppl}. Solid lines in Fig.~\ref{Fig2} are fit curves to the measurement data using the spin-diffusion model described above. Since laser-excitation initially creates a nonequilibrium state of the electrons that is not captured by the two-temperature model \cite{Tas1994}, we only fit decay and plateau of the measurement signal. Fit results are listed in Table~\ref{Tab1}, together with spin-diffusion lengths $\lambda_{\rm S} = \sqrt{D\tau_{\rm S}}$ that correspond to the spin-relaxation times determined. The errors listed in Table~\ref{Tab1} were determined from contours of constant variance $\sigma^2=2\sigma_{\rm fit}^2$ between model prediction and measurement data in the two-dimensional parameter space of $\tau_{\rm S}$ and $S_{\rm S}$, where $\sigma_{\rm fit}^2$ is the variance when $\tau_{\rm S}$ and $S_{\rm S}$ assume their fit values. Model parameters are summarized in the Supplemental Material, where we also demonstrate that the Faraday effect in the microscope objective does not contribute to our measurement signals, show exemplary measurements that demonstrate a sign change for negative magnetic fields, and present reference measurements on a Au/glass sample that show no measurement signal \cite{Suppl}. 

\par Though the FWHM of the time-correlation of pump and probe pulses is approximately 1.2~ps, the measurement signal does not rise before $t\approx 0$~ps. The delayed rise of the measurement signal cannot completely be explained by the finite diffusion time of spin and heat through the NM layer, which is considered in our model. The time delay between model and data during laser-excitation could correspond to the characteristic time of the scattering processes involved. This characteristic time can be estimated using the time-energy-correlation $\Delta t\propto h/\Delta E$, where $h$ is the Planck constant and $\Delta E$ is the interaction energy between electrons and magnons \cite{Carpene2015}. According to Ref.~\onlinecite{Rezende2014}, magnon frequencies in YIG at 300~K are of the order of 5~THz. This gives a characteristic time of the interfacial scattering process of $\Delta t\approx200$~fs, which is a factor 2-3 too small for explaining the delayed rise of the measurement signal. The remaining discrepancy could indicate that the two-temperature model fails in the sub-picosecond time scale, where the electron temperature is not well defined. 

\par Good agreement between model and measurement signal over the fit range for different layer thicknesses investigated and the finite measurement signal after electron-phonon thermalization in the NM layer support our assumption that the magnon temperature remains close to the phonon temperature. However, transfer of angular momentum across the NM/YIG interface is accompanied by energy transfer, which could lead to a reduction of the interfacial temperature difference between electrons and magnons. Therefore, we reanalyze the measurement data of the Au/YIG sample~I considering a two-temperature model of magnons and phonons in the YIG layer. Based on the fit result $\alpha \approx 10^8$~A~m$^{-2}$~K$^{-1}$ (compare Table~\ref{Tab1}), we estimate an electron-magnon thermal conductance across the NM/YIG interface of $G_{\rm em}=\alpha k_{\rm B}T/(2e)\approx10^6$~W~m$^{-2}$~K$^{-1}$. Assuming a magnetic heat capacity of YIG of $C_{\rm m}=1200$~J~m$^{-3}$~K$^{-1}$, theoretically calculated in Ref.~\onlinecite{Rezende2014}, we estimate a minimum magnon-phonon coupling constant of $g_{\rm mp}\approx 3\times10^{14}$~W~m$^{-3}$~K$^{-1}$, required to obtain fit results within the error bars of the results listed in Table~\ref{Tab1}. 

\par
\begin{figure}[h!]
	\centering
		\includegraphics[trim = 0mm 0mm 0mm 0mm,clip,width=8.7cm]{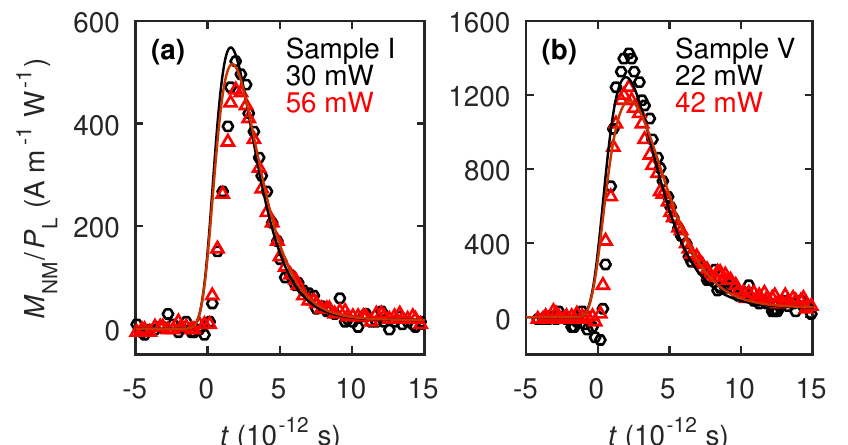}
	\caption{(Color online). \textbf{Fluence-dependent measurements.} TR-MOKE data (symbols) measured on sample~I (a) and sample~V (b) for different fluences as indicated. Solid lines were obtained from a simultaneous fitting of the spin-diffusion model to both data sets.}
	\label{Fig3}
\end{figure}

 Based on the temperature dependence of the electronic heat capacity, the temperature rise of electrons is nonlinear in the absorbed laser energy. If the measurement signal originates from spin accumulation driven by the temperature rise of electrons, it should also scale nonlinearly with absorbed laser energy. Figure~\ref{Fig3} shows TR-MOKE data normalized to the absorbed laser fluence for sample~I and sample~V. As expected, the normalized high-fluence data peaks below the normalized low-fluence data and shows a slightly delayed dynamics.  

\par
\begin{figure}[h!]
	\centering
		\includegraphics[trim = 0mm 0mm 0mm 0mm,clip,width=8.7cm]{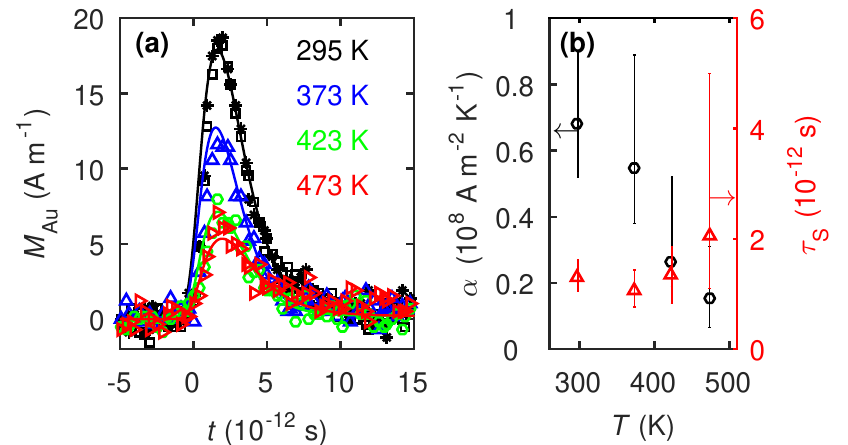}
	\caption{(Color online). \textbf{Temperature-dependent measurements.} (a) TR-MOKE data (symbols) measured on sample~I at different temperatures as indicated. Solid lines show fit curves obtained using the spin-diffusion model. (b) Fit results for $\alpha = g_{\uparrow\downarrow}e^2/hS_{\rm S}$ (left $y$-axis) and spin relaxation time $\tau_{\rm S}$ (right $y$-axis) as function of temperature.}
	\label{Fig4}
\end{figure}

 Based on the temperature dependence of the magnetization of YIG, the LSSE signal should decrease with increasing temperature and vanish at the Curie temperature. Figure~\ref{Fig4} shows measurement data at different ambient temperatures for sample~I in (a) and the temperature dependence of the respective fit results for $\alpha$ and $\tau_{\rm S}$ in (b). The measurement signals before and after heating are reversible [compare open squares and asteriks in (a)]. The size of the LSSE signal characterized by parameter $\alpha$ decreases monotonically towards the Curie temperature of YIG of approximately 550~K. The spin relaxation time does not show a significant temperature dependence within the errorbars of our measurements.

\par Weiler~\textit{et~al.}~report ISHE-based LSSE measurements on Pt/Au/YIG/GGG and Pt/Cu/YIG/GGG samples assuming interfacial LSSE, i.e., spin current driven by interfacial temperature difference between electrons and magnons \cite{Weiler2013}. In their model that is based on the theory of Ref.~\onlinecite{Xiao2010}, the parameter $\alpha$ is defined as
\begin{equation}
\label{EqW}
\widetilde\alpha  = \frac{g_{\uparrow\downarrow}\gamma e k_{\rm B}}{\pi M_{\rm S}V_{\rm a}},
\end{equation}
where $\gamma$ is the gyromagnetic ratio, $k_{\rm B}$ is the Boltzmann constant, $M_{\rm S}$ is the saturation magnetization, and $V_{\rm a}$ is the magnetic coherence volume. Weiler~\textit{et~al.} experimentally determine a spin-mixing conductance of Au/YIG and Cu/YIG interfaces of $g_{\uparrow\downarrow}\approx 4\times10^{18}$~m$^{-2}$. Using Eq.~(\ref{EqW}) with $M_{\rm S}=140$~kA~m$^{-1}$ and $V_{\rm a} = (1.3$~nm)$^3$ as reported by Weiler~\textit{et~al.} \cite{Weiler2013}, we obtain $\widetilde{\alpha}\approx 16\times10^8$~A~m$^{-2}$~K$^{-1}$, which is one order of magnitude larger than our results. Note that in addition to uncertainties in model parameters such as temperature difference between electrons and magnons and spin Hall angle, the measurements of Weiler\textit{et~al.}~include possible contributions from bulk LSSE. 

\par \textit{Conclusion.-} Using a novel method that is not based on the ISHE, we achieved LSSE measurements at the picosecond time-scale. Our experimental results corroborate LSSE theories that predict a spin current across the interface of a normal metal with a ferromagnetic insulator if magnons and electrons are out-of-equilibrium. We have isolated the interfacial LSSE and obtain a product of spin-mixing conductance and spin Seebeck coefficient of the order of $10^{8}$~A~m$^{-2}$~K$^{-1}$ for Au/YIG and Cu/YIG interfaces. Though our measurements indicate that the LSSE is active at the picosecond time scale, we find that the LSSE signal rises with a delay of 0.5~ps to 1~ps compared to our model prediction. To understand this delay, new LSSE theories are required that address the dynamics induced by sub-picosecond laser pulses. 

\begin{acknowledgments}
This work was carried out in part in the Frederick Seitz Materials Research Laboratory Central Research Facilities, University of Illinois. Financial supports by the Army Research Office under Contract No. W911NF-14-1-0016, by the German Research Foundation under DFG-Grant No. KI 1893/1-1 and KU 3271/1-1 (priority program Spin Caloric Transport, SPP 1538), and by the Department of Energy (DOE), Office of Science, Basic Energy Sciences, under Grants No. DE-SC0001304, are kindly acknowledged. We further thank Amit V.~Singh, Zhong Li and Arunava Gupta from Center for Materials for Information Technology (MINT), Tuscaloosa, Alabama, as well as G\"unter Reiss from Bielefeld University, Germany, for assistance during sample preparation and for making available the laboratory equipment.

\end{acknowledgments}


\newpage

\newpage

\newpage

\newpage

\newpage

\begin{table}[h!]
\caption{\label{Tab1} \textbf{Sample details and fit results.} Samples I, II, and III are from Ohio State University, samples IV, V, and VI are from University of Alabama and University of Bielefeld, Germany; $h$ denotes layer thickness; $\alpha = g_{\uparrow\downarrow}(e^2/h) S_{\rm S}$ (product of spin-mixing conductance per conductance quantum and spin Seebeck effect) and $\tau_{\rm S}$ (spin-relaxation time) are fit parameters; $\lambda_{\rm S}=\sqrt{D\tau_{\rm S}}$: spin-diffusion length. The errors were determined from contours of constant variance $\sigma^2=2\sigma_{\rm fit}^2$ between model prediction and measurement data in the two-dimensional parameter space of $\tau_{\rm S}$ and $S_{\rm S}$, where $\sigma_{\rm fit}^2$ is the variance when $\tau_{\rm S}$ and $S_{\rm S}$ assume their fit values.}
\vspace{0.5cm}
\begin{ruledtabular}
\begin{tabular}{rcccccc}
                                         & Sample~I & Sample~II & Sample~III & Sample~IV & Sample~V & Sample~VI \\
\hline
NM                                       & Au            & Au            & Cu            & Au            & Au            & Cu    \\
$h_{\rm NM}$~(nm)                        & 60            & 60            & 45            & 103           & 29            & 35    \\
$h_{\rm YIG}$~(nm)                       & 20            & 100           & 17            & 50            & 51            & 17    \\
$\alpha$ ($10^8$~A~m$^{-2}$~K$^{-1}$)    & 0.84$\pm$0.12 & 0.66$\pm$0.29 & 3.02$\pm$1.05 & 0.29$\pm$0.11 & 0.30$\pm$0.05 & 2.32$\pm$0.24  \\
$\tau_{\rm S}$~(ps)                      & 1.14$\pm$0.13 & 0.99$\pm$0.26 & 3.79$\pm$0.85 & 2.67$\pm$0.91 & 1.74$\pm$0.29 & 2.52$\pm$0.27   \\
$\lambda_{\rm S}$~(nm)                   & 86            & 81            & 172           & 130           & 111           & 161 \\
\end{tabular}
\end{ruledtabular}
\end{table}

\end{document}